\documentclass[prb,nofootinbib,preprintnumbers,tightenlines,twocolumn,superscriptaddress]{revtex4-2}
\usepackage{braket}
\usepackage{array}
\usepackage{dsfont}
\usepackage{amsmath}
\usepackage{pifont}
\usepackage[normalem]{ulem}
\usepackage{amssymb}
\usepackage{xcolor}
\usepackage{amsfonts}
\usepackage{mathtools}
\usepackage{dcolumn}
\usepackage{bm}
\usepackage{multirow}
\usepackage{hhline}
\usepackage[pdftex]{graphicx,hyperref}
\usepackage{leftidx}
\usepackage{color}
\usepackage{mathtools}
\usepackage{MnSymbol}
\usepackage[mathscr]{eucal}
\usepackage[german,english]{babel}
\usepackage[capitalize]{cleveref}

\usepackage[utf8]{inputenc}
\usepackage{natbib}
\usepackage{comment}

\usepackage{soul}

\begin{document}
	
	
	\title{Dynamical Effects from Anomaly: Modified Electrodynamics in Weyl Semimetal}
	
	\author{Xuzhe Ying}
	\affiliation{Department of Physics and Astronomy, University of Waterloo, Waterloo, ON, N2L 3G1, Canada}
	\affiliation{Perimeter Institute for Theoretical Physics, Waterloo, ON, N2L 2Y5, Canada}
	
	\author{A. A. Burkov}
	\affiliation{Department of Physics and Astronomy, University of Waterloo, Waterloo, ON, N2L 3G1, Canada}
	\affiliation{Perimeter Institute for Theoretical Physics, Waterloo, ON, N2L 2Y5, Canada}
	
	\author{Chong Wang}
	\affiliation{Perimeter Institute for Theoretical Physics, Waterloo, ON, N2L 2Y5, Canada}

	\begin{abstract}
		We discuss the modified quantum electrodynamics from a time-reversal-breaking Weyl semimetal coupled with a $U(1)$ gauge (electromagnetic) field. A key role is played by the soft dispersion of the photons in a particular direction, say $\hat{z}$, due to the Hall conductivity of the Weyl semimetal. 
		Due to the soft photon, the fermion velocity in $\hat{z}$ is logarithmically reduced under renormalization group flow, together with the fine structure constant. Meanwhile, fermions acquire a finite lifetime from spontaneous emission of the soft photon, namely the Cherenkov radiation. At low energy $E$, the inverse of the fermion lifetime scales as $\tau^{-1}\sim E/{\rm PolyLog}(E)$. Therefore, even though fermion quasiparticles are eventually well-defined at very low energy, over a wide intermediate energy window the Weyl semimetal behaves like a marginal Fermi liquid. Phenomenologically, our results are more relevant for \textit{emergent} Weyl semimetals, where the fermions and photons all emerge from strongly correlated lattice systems.   Possible experimental implications are discussed.
		\begin{description}
			\item[PACS numbers]
		\end{description}
	\end{abstract}
	
	\pacs{1111}
	\maketitle
	
	\section{Introduction}
	
	Weyl fermions, since their original proposal, have been widely studied due to the chiral nature [\onlinecite{weinberg1995quantum,peskin2018introduction,volovik2003universe,nielsen1983adler,nielsen1981absenceI,nielsen1981absenceII,PhysRevLett.107.127205,PhysRevX.5.031013,HasanWeylDiscovery}]. In the recent decades, much focus has been put on the condensed matter realization, namely the Weyl semimetal (WSM) [\onlinecite{PhysRevLett.107.127205,PhysRevX.5.031013,HasanWeylDiscovery,RevModPhys.90.015001}]. In Weyl semimetals, due to the separation of Weyl fermions in momentum space, various intriguing phenomena have been observed, e.g., Fermi arc [\onlinecite{PhysRevX.5.031013},\onlinecite{HasanWeylDiscovery}], anomalous Hall effect [\onlinecite{PhysRevLett.107.127205},\onlinecite{Zyuzin12-1}], quantized circular photogalvanic effect [\onlinecite{de2017quantized}], etc. The dynamical properties of WSM also attract much attention [\onlinecite{PhysRevLett.116.116803,PhysRevLett.126.185303,PhysRevB.87.161107,PhysRevB.90.035126}]. While a magnetic Weyl semimetal is typically subject to weak interaction of particular form (short-ranged or Coulomb), an emergent WSM is a more versatile playground for studying interaction effects, e.g., topological orders in three dimension \onlinecite{PhysRevB.98.201111,PhysRevLett.124.096603,PhysRevB.101.235168}], generalizations of the standard QED \onlinecite{PhysRevD.41.1231}, etc.
	
	 An emergent WSM is a strongly interacting lattice system of spins or electrons, of which the low energy effective theory is described by an emergent $U(1)$ gauge field (or some $\mathbb{Z}_m$ descendent) coupled to a WSM formed by emergent fermions. In spin liquid terminology, these are $U(1)$ (or $\mathbb{Z}_m$) spin liquids with spinon Weyl semi-metals. Unlike in ordinary Weyl semimetals, the emergent Weyl fermions could naturally have velocity close to that of the $U(1)$ gauge field, and the gauge coupling strength (fine structure constant) does not have to be small at a given energy scale.  The possibility of an emergent WSM phase has been demonstrated in Ref.~[\onlinecite{PhysRevB.98.201111,PhysRevResearch.1.033029,PhysRevB.94.155136,PhysRevLett.113.136402,10.21468/SciPostPhys.8.2.031}].  
	 The emergent WSM phase was further proposed to be the parent state of topological orders in three dimension [\onlinecite{PhysRevB.98.201111,PhysRevLett.124.096603,PhysRevB.101.235168}]. While the descendent topologically ordered phases are stable by the formation of many-body gap, the properties of the emergent WSM phase itself is largely studied at the mean-field level. In particular, the dynamical consequences of gauge fluctuations in emergent WSM remain unexplored. 
	
	
	In this work, we focus on the case with the emergent U(1) gauge field, also referred to as emergent electromagnetic (EM) field. One important notion in studying WSM phase is the unquantized anomaly, which guarantees the gaplessness of WSM [\onlinecite{Zyuzin12-1,Grushin12,PhysRevResearch.3.043067,PhysRevB.104.235113}]. When an EM field emerges, the dynamical aspect of the anomalies is also an important piece of information. The unquantized anomaly appears as a Chern-Simons-like action in $3+1$D [\onlinecite{Zyuzin12-1,Grushin12,PhysRevResearch.3.043067,PhysRevB.104.235113}]. Together with the Maxwell action, the modified electrodynamics is usually referred to as Carroll-Field-Jackiw electrodynamics [\onlinecite{PhysRevD.41.1231}]. In the modified electrodynamics, the physical polarization of propagating photons is different from those in the vacuum. Another important feature is the anisotropy. In particular, one of the photon modes becomes soft in a particular direction [\onlinecite{PhysRevD.41.1231}]. Emergent photons with similar features were also found in the coupled layers of Laughlin states [\onlinecite{PhysRevB.79.235315}].
	
    In this article, we study the interplay between the fermionic degrees of freedom and the modified electrodynamics in emergent WSM. The situation under consideration is really a simple, non-Lorentz-symmetric generalization of textbook quantum electrodynamics (QED), however with unconventional outcomes. Indeed, we will show that due to the interaction with the soft photons, the emergent WSM represents an unconventional quantum liquid. 
 	
    More specifically, the presence of soft photons significantly influences the low-energy properties of the fermions. There are two major results. First, the fermion dispersion is strongly dressed by the photons. Namely, the fermion's velocity in the soft photon direction is reduced to zero under the rernormalization group (RG) flow. Besides, the system flows to a non-interacting limit under RG. Second, fermions can spontaneously emit photons. As a result, the fermions acquire a finite lifetime, due to the Cherenkov radiation of the soft photons, that is inversely proportional to the fine structure constant and the fermion's energy. The two effects just mentioned make the emergent WSM significantly different from the free WSM or the standard QED. Indeed, over a wide energy window, the emergent WSM behaves like a marginal Fermi liquid due to the finite lifetime [\onlinecite{PhysRevLett.63.1996}]. Meanwhile, the IR behavior shows an asymptotic two dimensional character and is well-controlled under the RG flow. Hence, the emergent WSM represents an unconventional type of quantum liquid. Lastly, we propose that the reported feature of a emergent WSM can be observed from the specific heat measurement at low temperature [\onlinecite{PhysRevLett.102.176404},\onlinecite{PhysRevLett.118.246601}].
	
	The rest of the article is organized as follows. Section.~\ref{sec:WSM&ED&Intuition} reviews the low energy description of a Weyl semimetal as well as the modified electrodynamics. A physical picture is developed for the dynamical effects. Section.~\ref{sec:LoopCalculations} is devoted to quantum mechanical one-loop diagram calculations, from which the RG flow and the fermion lifetime can be obtained. Section.~\ref{sec:Conclusion} concludes the article and discusses the possible experimental implications.
	
	\section{(Emergent) Weyl Semimetal and Modified Electrodynamics}
	\label{sec:WSM&ED&Intuition}
	
	In this section, we review the low energy description of the fermions and the modified electrodynamics in emergent Weyl semimetal (WSM) at mean field level. A qualitative picture for the dynamical effects is provided.
	
\begin{figure}[t]
	    \centering
	    \includegraphics[scale=0.25]{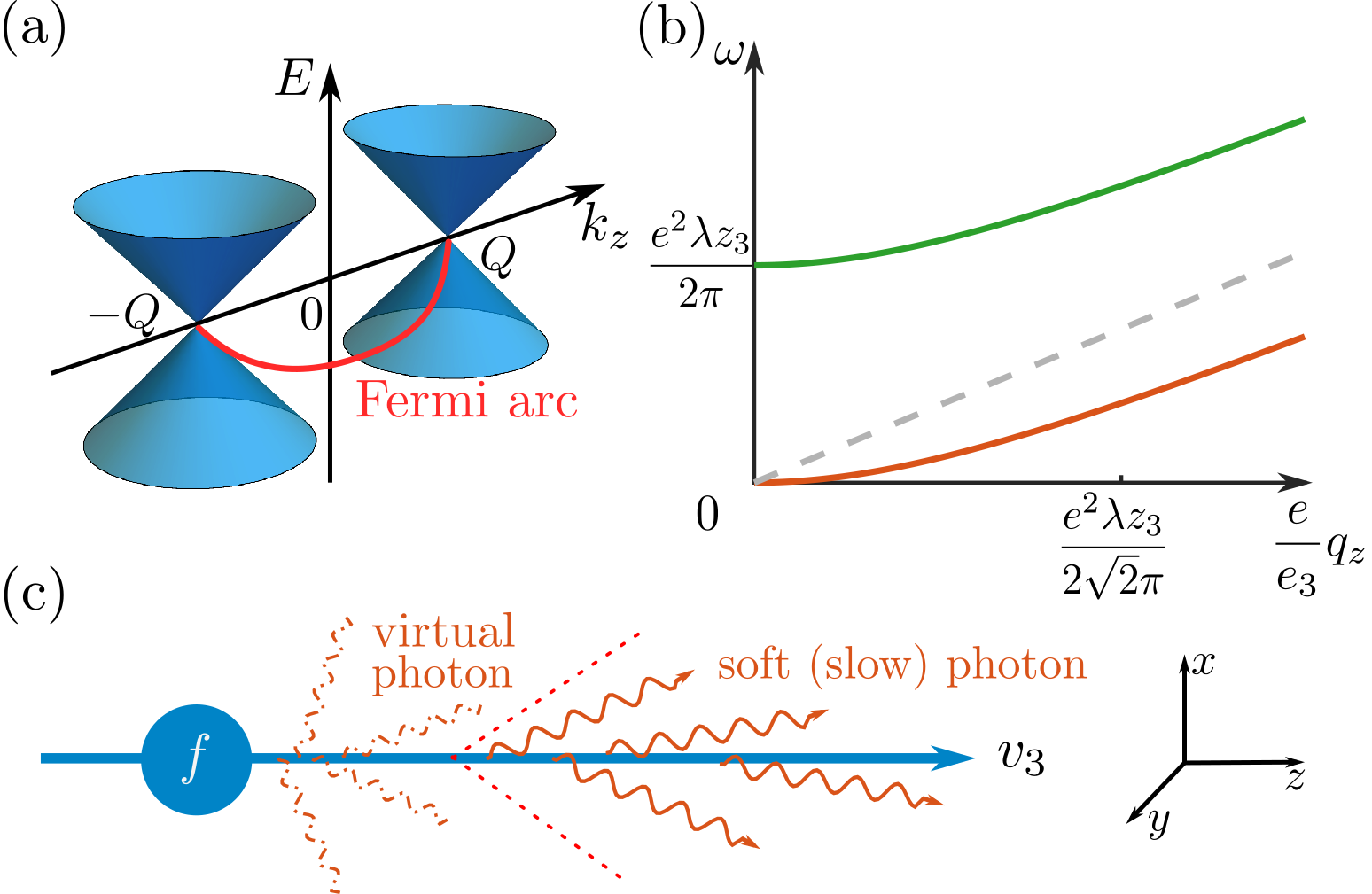}
	    \caption{Low energy dispersions for (a) free WSM; (b) photon in the modified electrodynamics with $q_x=q_y=0$. (a) Around two Weyl nodes, fermions have linear dispersion. The two Weyl nodes are separated in momentum space by $2Q=\lambda z_3$ in $z$-direction. On the sample boundary, there is Fermi arc connecting two Weyl nodes, an indication of anomalous Hall effect. (b) There are two photon modes. The gapped mode (green) has a gap given by $\frac{e^2\lambda z_3}{2\pi}$. The other mode is soft (red), with a quadratic dispersion, Eq.~(\ref{eq:SoftPhotonDispersion}). The grey dahsed line corresponds to $\omega=\frac{e}{e_3}q_z$, namely the photon dispersion when $\lambda=0$. (c) Cartoon of fermions moving in a soft electromagentic environment. Fast moving fermions constantly interact with virtual photons (red dashed-dot) and spontaneously emit soft photons (red solid). The emitted soft photons are roughly confined around the $z$-axis indicated by the red dashed lines.}
	    \label{fig:MeanFieldDispersion}
	\end{figure}

    The mean field description of the emergent WSM starts with a parton decomposition, which formally corresponds to fractionalizing the electron annihilation operator $c$ into a neutral fermion (spinon, $\psi$) and a charged boson (chargeon, $e^{i\theta_c}$): $c= e^{i\theta_{\text{c}}}\psi$ [\onlinecite{LeeNagaosaWen2006}]. The local $U(1)$ gauge ambiguity of this decomposition
    \begin{equation}
       \psi\to e^{i\alpha}\psi, \hspace{10pt} e^{i\theta_c}\to e^{-i\alpha}e^{i\theta_c},
    \end{equation}
    dictates the necessity of the emergence of a dynamical $U(1)$ gauge field [\onlinecite{Wen_QFT}]. In other words, there is a dynamical $U(1)$ gauge field that couples to both the  the spinon and the chargeon with gauge charge $q=\pm1$. We then consider mean field states in which the chargeons $e^{i\theta_c}$ are gapped (and therefore can be integrated out at low energy), while the spinons $\psi$ form a Weyl semimetal band structure, with two Weyl cones of the opposite chirality separated in momentum space by $2\boldsymbol{Q}$ (see Fig.~\ref{fig:MeanFieldDispersion}~(a)). 
	
	At low energy, the fermionic excitations of a Weyl semimetal is effectively described by a Dirac Lagrangian [\onlinecite{Zyuzin12-1}]:
	\begin{equation}
	    \mathcal{L}_{\text{f}}=\bar{\psi}\ \mathcal{V}^{\mu}i\partial_{\mu}\psi,\ \text{with }\mathcal{V}^{\mu}=[\gamma^0,\gamma^1,\gamma^2,v_3\gamma^3]
	    \label{eq:DiracLagrangian}
	\end{equation}
	where $\gamma^{0,1,2,3}$ are the usual $4\times 4$ gamma matrices. Here, we assumed the velocity in the $xy$-plane to be one while the velocity in the $z$-direction being different, namely $v_3$. The energy of the fermions is $E_{\pm}(\boldsymbol{k})=\pm E(\boldsymbol{k})=\pm\sqrt{k_x^2+k_y^2+v_3^2k_z^2}$. We should comment that in this article, we interchangeably use $\mu=(0,1,2,3)$ or $\mu=(t,x,y,z)$ to indicate the time and space directions. The former is algebraically convenient, while the latter more descriptive and intuitive.
	
	One should notice that the simplicity in Eq.~(\ref{eq:DiracLagrangian}) is deceptive. Indeed, when a lattice model of WSM is considered, the two Weyl components of a Dirac fermion are separated in Brillouin zone [\onlinecite{PhysRevLett.107.127205}], as shown in Fig~\ref{fig:MeanFieldDispersion}(a). As a result, there will be chiral edge states for a finite size system. The intersection of chiral edge state and the Fermi level is the Fermi arc, which is a hallmark of WSM [\onlinecite{PhysRevX.5.031013},\onlinecite{HasanWeylDiscovery}]. In terms of transport, the features just mentioned implies that WSM shows an anomalous Hall conductivity [\onlinecite{PhysRevLett.107.127205},\onlinecite{Zyuzin12-1},\onlinecite{PhysRevLett.113.187202},\onlinecite{RevModPhys.90.015001}]. The anomaly argument requires the value of the Hall conductivity to be given by $2\boldsymbol{Q}$, even though high-energy fermions may be strongly interacting with no well-defined quasi-particles [\onlinecite{PhysRevLett.124.096603},\onlinecite{PhysRevResearch.3.043067}]. 
	
	
	Due to the Hall effect, the Lagrangian for the emergent U(1) gauge field has a Chern-Simons(CS)-like term, in addition to the Maxwell term [\onlinecite{PhysRevD.41.1231}]: 
	\begin{subequations}
	    \begin{align}
	        &\mathcal{L}_{\text{CS-like}}=-\frac{\lambda z_3}{4\pi}\left(\phi B_z-a_xE_y+a_yE_x\right) \label{eq:CSLikeTerm}\\
	        &\mathcal{L}_{\text{M}}=\frac{1}{2e^2}\left(E_x^2+E_y^2-B_z^2\right)+\frac{1}{2e_3^2}\left(E_z^2-B_x^2-B_y^2\right) \label{eq:MaxwellTerm}
	    \end{align}
	    \label{eq:CSLike_Maxwell_Lagrangian}
	\end{subequations}
where $\phi=a_0$ and $\boldsymbol{a}$ is the scalar and vector potential of the gauge field; $\boldsymbol{E}$ and $\boldsymbol{B}$ are the emergent electric and magnetic field strength respectively. 
	
	The first line, Eq.~(\ref{eq:CSLikeTerm}), is the CS-like term and is responsible for the Hall effect. In this work, $z_3$ is essentially the reciprocal lattice constant along $z$-direction. As shown in Fig.~\ref{fig:MeanFieldDispersion}(a), $2Q =\lambda z_3$ measures the separation between Weyl points in momentum space. We should mention that to emphasize the anomaly aspect, the CS-like term is sometimes compactly written as $\mathcal{L}_{\text{CS-like}}\sim z\wedge a\wedge da$ in terms of differential forms and the translation gauge field $z_{\mu}=(0,0,0,z_3)$ [\onlinecite{PhysRevResearch.3.043067},\onlinecite{PhysRevB.104.235113}]. Indeed, the CS-like term dictates the mixed unquantized anomaly of translation and U(1) gauge symmetry. The mixed anomaly is a manifestation of the chiral anomaly. As we will point out, the dynamical effect is also important. Namely, the CS-like term significantly alters the photon dispersion, which in turn influences the properties of the fermions as well as the infrared description of an emergent WSM.
	
	The second line, Eq.~(\ref{eq:MaxwellTerm}), is the Maxwell Lagrangian. Here, two coupling constants $e$ and $e_3$ are introduced. The two coupling constants introduce some anisotropy in the Maxwell term, as will be clear below in the photon dispersions.
	

	The photon dispersions can be straightforwardly found based on the Euler-Lagrange equation for the $U(1)$ gauge field Lagrangian, Eq.~(\ref{eq:CSLike_Maxwell_Lagrangian}). There are two propagating photon modes with the following dispersion relation [\onlinecite{PhysRevD.41.1231}]:
	\begin{equation}
	    \omega^2_{\pm}=q_x^2+q_y^2+\frac{e^2}{e_3^2}q_z^2+e^4\frac{\lambda^2z_3^2}{8\pi^2}\pm e^4\frac{\lambda^2z_3^2}{8\pi^2}\sqrt{1+2\frac{8\pi^2}{e^4\lambda^2z_3^2}\frac{e^2}{e_3^2}q_z^2}
	\end{equation}
	When the CS-like term vanishes, $\lambda=0$, the photons retain a linear dispersion as usual, with an anisotropy in the speed of light: the speed of light in $xy$-plane is unity, while in the $z$-direction being $c_z=\frac{e}{e_3}$. 
	
	The presence of CS-like term, $\lambda\neq 0$, significantly alters the photon dispersion. As plotted in Fig.~\ref{fig:MeanFieldDispersion}(b), there are two photon modes: a gapped mode as well as a gapless soft mode. The gapped photon mode has a gap $e^2\frac{\lambda z_3}{2\pi}$. Notice that the gap only depends on the coupling $e^2$, not $e_3^2$. This mode mimics the gapped mode of Maxwell-Chern-Simons theory in $2+1$D.
	
	The other photon mode is gapless and quite soft. At low energy, the soft photon takes the following dispersion:
	\begin{equation}
	    \omega_-^2\approx q_x^2+q_y^2+\frac{4\pi^2}{e_3^4\lambda^2z_3^2}q_z^4
	    \label{eq:SoftPhotonDispersion}
	\end{equation}
	Namely, the soft photon dispersion is extremely anisotropic due to the CS-like term, Eq.~(\ref{eq:CSLikeTerm}). When moving along the $z$-direction, the photon has a quadratic dispersion with an effective inertia mass $m_{\text{s}}\sim e_3^2\lambda z_3$. Notice that only $e_3^2$ (not $e^2$) enters the low energy dispersion at leading order. Such soft nature will play a central role in our subsequent analysis.
	
	Finally we consider the fermions and the photons to be minimally coupled:
	\begin{equation}
	    \mathcal{L}_{\text{c}}=\bar{\psi}\ \mathcal{V}^{\mu}a_{\mu}\psi.
	    \label{eq:MinimalCoupling}
	\end{equation}
	Eq.~(\ref{eq:DiracLagrangian}), Eq.~(\ref{eq:CSLike_Maxwell_Lagrangian}) and Eq.~(\ref{eq:MinimalCoupling}) form our full Lagrangian. Such Lagrangian captures all the essential infra-red (IR) features of an emergent WSM, while containing only a limited number of parameters.
	
	The coupling between the fermions and the photons puts some dynamical constraint on the kinematics. In the absence of CS-like term, Eq.~(\ref{eq:CSLikeTerm}), the fermion velocity and the speed of light renormalizes to be identical in all directions [\onlinecite{PhysRevD.83.105027}]. The same holds true for the velocities in $xy$-plane when the CS-like term is present. The renormalization effect is more relevant when the gauge field is emergent, in which case the typical velocities of photons and fermions are actually comparable [\onlinecite{Wen_QFT}]. Based on the considerations just mentioned, we have assumed the velocities of both fermions and photons in $xy$-plane to be unity. More drastic effect occurs for the motion in $z$-direction.
	
	A cartoon of fermions moving in $z$-direction is depicted in Fig.~\ref{fig:MeanFieldDispersion}(c). In emergent WSM, the linearly dispersive fermions and the soft photon coexist and are coupled. In the $z$-direction, fermions move much faster than the photons at low energy. There are two major effects following from this observation.
	
	One effect is related to the photon dressing of fermion's dispersion. Namely, fermions constantly emit and absorb virtual photons, Fig.~\ref{fig:MeanFieldDispersion}(c). Pictorially, the fermions constantly ``hit'', in particular, the soft photons. Because of the slow velocity, the soft photons serve as an impedance of fermion's motion in $z$-direction. Therefore, one would expect that the fermion is dressed by the soft photons and the velocity in $z$-direction being reduced.
	
	Another effect is that fermions can radiate photons spontaneously. As schematically shown in Fig.~\ref{fig:MeanFieldDispersion}(c), the radiated photons are all soft, with momentum roughly confined in a cone structure around $z$-axis. This is exactly the picture of Cherenkov radiation in classical electrodynamics [\onlinecite{schwinger1998classical},\onlinecite{PhysRevD.98.114026}]. Thus, the fermions will acquire a finite lifetime. As we will show later, the inverse of the fermion's lifetime is proportional to its energy and the fine structure constant. 
	
	In conclusion, we expect that the soft photon renormalizes the fermion velocity in $z$-direction and gives fermion a finite lifetime. Such expectation is supported by the quantum mechanical 1-loop calculations as detailed in Section.~\ref{sec:LoopCalculations}. In the end, the emergent WSM represents an unconventional quantum liquid, whose infrared properties differ significantly from the non-interacting WSM or the standard QED. The unconventional features are the result of the CS-like term, namely the mixed unquantized anomaly of WSM.
	
	\section{Implications from 1-loop Diagrams}
	\label{sec:LoopCalculations}

	\begin{figure}[t]
	    \centering
	    \includegraphics[scale = 0.35]{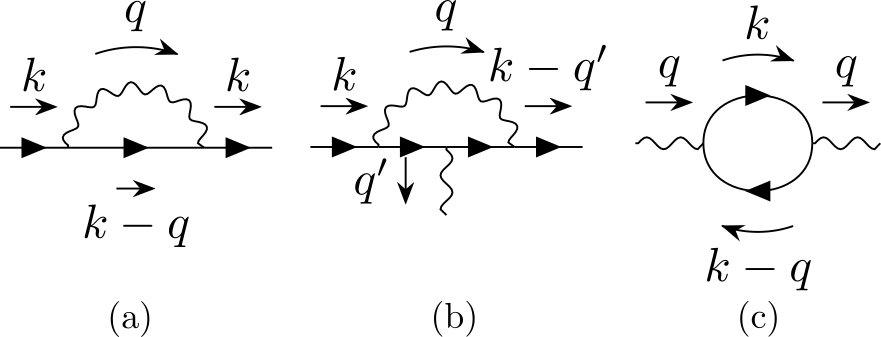}
	    \caption{One loop diagrams: (a) Fermion self-energy; (b) Interaction vertex correction; (c) Polarization operator.}
	    \label{fig:OneLoopDiagrams}
	\end{figure}
	
	In this section, we will show that the qualitative picture developed in Section.~\ref{sec:WSM&ED&Intuition} is supported by quantum mechanical calculation. In particular, we study the 1-loop diagrams in Fig.~\ref{fig:OneLoopDiagrams} as in the standard QED [\onlinecite{weinberg1995quantum},\onlinecite{peskin2018introduction}].
	
	We first investigate the full set of the 1-loop diagrams, Fig.~\ref{fig:OneLoopDiagrams}. We will show that both the fermion's velocity, $v_3$, and the Maxwell speed of light, $c_z=e/e_3$, in $z$-direction renormalizes to zero. At the same time, the electromagnetic coupling constants renormalize to zero. Namely, in the IR, the photon dressed fermions are non-interacting and shows an asymptotic 2D dispersion.
	
	Then, we study the imaginary part of the fermion self-energy, Fig.~\ref{fig:OneLoopDiagrams}(a). We will show that the inverse of the fermion's lifetime (and equivalently the one-photon emission rate) is proportional to the fermion's energy. At weak (and even moderate) coupling, the proportionality vanishes logarithmically in the IR limit due to the RG effect. The associated photon radiation process is the quantum version of Cherenkov radiation for the modified electrodynamics of Eq.~(\ref{eq:CSLike_Maxwell_Lagrangian}) [\onlinecite{PhysRevD.98.114026}]. Besides, we should mention that the fermion lifetime is also sensitive to the direction of fermion's momentum. Therefore, the emergent WSM represents an unconventional quantum liquid. Over a wide energy range, it behaves like a marginal Fermi liquid, yet with a well-controlled IR behavior.  
	
	Before proceeding, it's worthwhile to discuss the typical energy scales. There are three typical energy scales in the current setup. One is $\Lambda_{\text{f}}\sim v_3\lambda z_3$. Below $\Lambda_{\text{f}}$, the linear approximation of fermion dispersion, Eq.~(\ref{eq:DiracLagrangian}), applies. Second is the photon gap, $m_{\text{p}}\sim e^2\lambda z_3$, below which the gapped photon mode decouples from the rest of the fields. The last one is slightly subtler: at low enough energy, the photon mode $a_{x,y}$, with quadratic dispersion in $\hat{z}$, $\omega_{ph}\sim q_z^2/2m_s$ with inertia mass $m_{\text{s}}\sim e_3^2\lambda z_3$, and the fermions, with linear dispersion $\omega_{f}\sim v_3k_3$, will not be able to couple efficiently due to a large energy mismatch when $q_3$ is comparable to $k_3$. This kinematic constraint leads to the dynamical decoupling of the quadratic-dispersing photon mode with other fields in the theory, which happens when the energy scale is below $m_sv_3^3\sim v_3^2e_3^2\lambda z_3$. Below we will work with energy scale $\Lambda\ll\text{Min}\left\{\Lambda_{\text{f}},m_{\text{p}},m_{\text{s}}v_3^2\right\}$.
	
	With such small energy scale (and the corresponding long wavelength), the photon propagator is primarily given by:
	\begin{equation}
	    G^{\text{a}}_{\mu\nu}(q)=e_3^2\frac{1}{q_0^2-q_1^2-q_2^2}\frac{z_{\mu}z_{\nu}}{z_3^2}
	    \label{eq:GaugeFieldPropagator}
	\end{equation}
	where $z_{\mu}=(0,0,0,z_3)$. The corrections are on the order of $\mathcal{O}(q/m_{\text{p,s}})\ll 1$. There are also dependence on the gauge fixing parameter, which may not be small in magnitude. Nevertheless, the gauge dependence does not enter any physical properties, as generally expected from the gauge invariance principle.
	
	One last ingredient is the fermion propagator:
	\begin{equation}
	    G_{\text{f}}(k)=\frac{\mathcal{V}^{\mu}k_{\mu}}{k_0^2-k_1^2-k_2^2-v_3^2k_3^2}
	    \label{eq:FermionPropagator}
	\end{equation}
	Notice that both the gauge field and the fermion propagator shows some anisotropy. The anisotropy in the gauge field propagator is more important. Indeed, at low energy, only one pole is present in Eq.~(\ref{eq:GaugeFieldPropagator}). This pole exactly corresponds to the soft photon at low energy and long wavelength. (Notice that $q_z$ dependence of the soft photon dispersion in Eq.~(\ref{eq:SoftPhotonDispersion}) is now subleading and neglected.) Therefore, the unconventional behaviors of an emergent WSM to be reported in this section are indeed due to the soft photon.

	\subsection{Fermion Velocity Renormalization and the RG Equations}
	
	\label{sec:RGAnalysis}
	
	For weak interaction, the fermionic excitations are presumably under-damped. Then, we can ask the question of how fermion band structure renormalizes. Along $z$-direction, the fast moving fermions are dressed by the slow moving photons due to the interaction. Therefore, a reduction in the fermion velocity, $v_3$, is expected from the RG analysis.
	
	To be more specific, we aim at the renormalization of fermion and photon's dispersion as well as the coupling strength between them. Namely, we would like to obtain the RG flow of $v_3$, $e$ and $e_3$. To achieve this goal, the following bare scaling dimension is considered:
	\begin{equation}
	    [k_{\mu}]=[q_{\mu}]=1,\ [\bar{\psi}]=[\psi]=\frac{3}{2},\ [a_{\mu}]=1
	    \label{eq:BareScaling}
	\end{equation}
	Under this bare scaling, the CS-like term, Eq.~(\ref{eq:CSLikeTerm}), is relevant. Meanwhile, all the other terms in the Lagrangian, Eq.~(\ref{eq:DiracLagrangian}), Eq.~(\ref{eq:MaxwellTerm}) and Eq.~(\ref{eq:MinimalCoupling}), are marginal. The relevance of the CS-like term implies that all the three typical scales, $\Lambda_{\text{f}}$, $m_{\text{p}}$ and $m_{\text{s}}v_3^2$, renormalizes to infinity. This is indeed the case when we consider the full RG equation below. The relevance of the three typical energy scales is also reflected in the gauge field propagator. As shown in Eq.~(\ref{eq:GaugeFieldPropagator}), at low energy, only one pole at $q_0^2=q_1^2+q_2^2$ is present in the leading order term of the gauge field propagator.
	
	The fermion velocity renormalization can be obtained from either the fermion self-energy, $\Sigma(k)$ in Fig.~\ref{fig:OneLoopDiagrams}(a), or the vertex correction, $\Gamma^{\mu}(k,q^{\prime})$ in Fig.~\ref{fig:OneLoopDiagrams}(b). The two diagrams are related by the Ward identity. In particular, we focus on the following quantity [\onlinecite{weinberg1995quantum},\onlinecite{peskin2018introduction}]:
	\begin{equation}
	    \Gamma_{\text{1-loop}}^{\mu}(k,q^{\prime}=0)|_{k=0}=-\partial_{k_{\mu}}\Sigma(k)|_{k=0}
	\end{equation}
	Combining with the bare interaction vertex, Eq.~(\ref{eq:MinimalCoupling}), the full interaction vertex is given by:
	\begin{equation}
	    \Gamma_{\text{full}}^{\mu}(k,q^{\prime}=0)|_{k=0}=Z^{-1}_{\text{f}}\left[\gamma^0,\gamma^1,\gamma^2,\tilde{v}_3\gamma^3\right]
	\end{equation}
	where all the gauge dependence enters the fermion field renormalization factor $Z_{\text{f}}$. As a result, the effective velocity, $\tilde{v}_3$, is gauge independent:
	\begin{equation}
	    \tilde{v}_3=v_3-\frac{e_3^2}{6\pi^2}\frac{v_3^3}{|v_3|}\ln \frac{\Lambda}{\mu_{\text{IR}}}
	    \label{eq:EffVelocity}
	\end{equation}
	where $\Lambda$ and $\mu_{\text{IR}}$ are the UV and IR cutoff in the loop momentum integral. This result suggests that fermion velocity, $v_3$, is reduced by the photon dressing.
	
    The renormalization of the coupling constants $e$ and $e_3$ can be obtained from the vacuum polarization, $\Pi^{\mu\nu}(q)$ as in Fig.~\ref{fig:OneLoopDiagrams}(c). Technically, the vacuum polarization is evaluated with the dimensional regularization scheme, in order to maintain gauge invariance [\onlinecite{weinberg1995quantum},\onlinecite{peskin2018introduction}]. Assuming the spacetime dimension as $d=4-\epsilon$, the diagram of Fig.~\ref{fig:OneLoopDiagrams}(c) contains a factor of $\Pi^{\mu\nu}(q)\propto\Lambda^{\epsilon}(\frac{2}{\epsilon}-\ln\Delta^2(q))\sim\ln\frac{\Lambda^2}{\Delta^2(q)}$, which is reduced to a logarithmically divergent factor by minimal subtraction. Here, $\Delta^2(q)=q_0^2-q_1^2-q_2^2-v_3^2q_3^2$.
    
    Physically, the vacuum polarization operator, Fig.~\ref{fig:OneLoopDiagrams}(c), gives a correction to the bare Maxwell term, Eq.~(\ref{eq:MaxwellTerm}). The resultant Lagrangian is of the same form as Eq.~(\ref{eq:MaxwellTerm}), with the effective coupling constants:
    \begin{equation}
        \begin{split}
           &\frac{1}{\tilde{e}^2}=\frac{1}{e^2}+\frac{N_{\text{f}}}{12\pi^2}\frac{1}{|v_3|}\ln\frac{\Lambda^2}{\Delta^2(q)} \\
           &\frac{1}{\tilde{e}_3^2}=\frac{1}{e_3^2}+\frac{N_{\text{f}}}{12\pi^2}|v_3|\ln\frac{\Lambda^2}{\Delta^2(q)}
        \end{split}
        \label{eq:EffCoupling}
    \end{equation}
    where $N_{\text{f}}$ is the number of pairs of the Weyl points.
	
	Based on the bare scaling dimension Eq.~(\ref{eq:BareScaling}) and the results of Eq.~(\ref{eq:EffVelocity}) and Eq.~(\ref{eq:EffCoupling}), we obtain the main results, namely the RG equations:
	\begin{subequations}
	    \begin{align}
	        &\frac{dz_3}{dl}=z_3 \label{eq:z3RG}\\
	        &\frac{d v_3}{dl}=-\frac{e_3^2}{6\pi^2}\frac{v_3^3}{|v_3|} \label{eq:v3RG}\\
	        &\frac{de^2}{dl}=-N_{\text{f}}\ \frac{e^4}{6\pi^2}\frac{1}{|v_3|},\ \  \frac{de_3^2}{dl}=-N_{\text{f}}\ \frac{e_3^4}{6\pi^2}|v_3|
	    \end{align}
	    \label{eq:RGequations}
	\end{subequations}
	where the RG process is defined through the change in the UV scale as $\Lambda\rightarrow\Lambda\exp[-dl]$ with $dl>0$. The RG equation for the reciprocal lattice constant $z_3$, Eq.~(\ref{eq:z3RG}), follows from the bare scaling Eq.~(\ref{eq:BareScaling}).
	
	\begin{figure}[t]
	    \centering
	    \includegraphics[scale = 0.325]{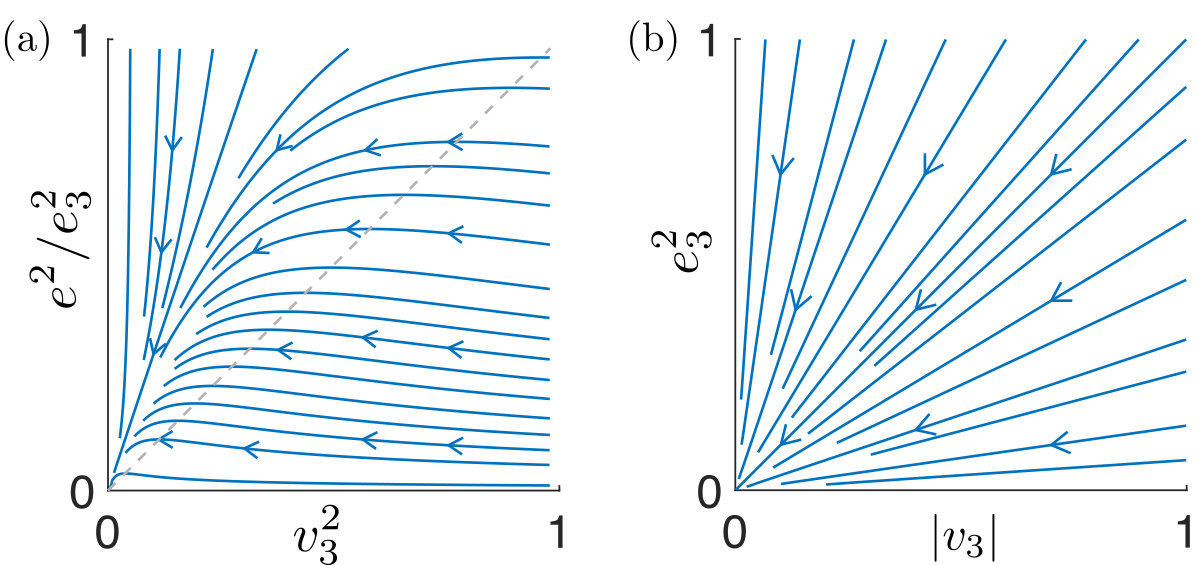}
	    \caption{Renormalization group flow with $N_{\text{f}}=1$ pair of Weyl fermions.}
	    \label{fig:RGFlow}
	\end{figure}
	
	Fig.~\ref{fig:RGFlow} plots the RG flow of a few typical parameters. Fig.~\ref{fig:RGFlow}(a) shows the RG flow of fermion velocity, $v_3$, and the speed of light (as defined solely from Maxwell term), $c_z=e/e_3$. While the fermion velocity always flows to a smaller value, there is a slow increase in $c_z$ when the fermion velocity is large $v_3>c_z$. Nevertheless, both velocities flows to zero eventually. Fig.~\ref{fig:RGFlow}(b) shows that in addition to the reduction of the velocities in $z$-direction, the coupling constant $e_3$ also flows to zero. From the RG flow, we conclude that in the deep IR limit, the emergent WSM coupled with dynamical U(1) gauge field shows asymptotic two-dimensional dispersion and is free of interaction. The weakly coupled IR limit also justifies our perturbative loop expansion.
	
	We should comment on the flow of the three typical energy scales, $\Lambda_{\text{f}}$, $m_{\text{p}}$ and $v_3^2m_{\text{s}}$. Notice that those scales are of the form $\{v_3,e^2,v_3^2e_3^2\}\times z_3$. From the RG equation, the fermion velocity and the coupling constants renormalizes to zero as $1/{\rm PolyLog}(\Lambda)$, as suggested by Eq.~(\ref{eq:EffVelocity}) and Eq.~(\ref{eq:EffCoupling}). Meanwhile, the reciprocal lattice constant $z_3$ renormalizes to infinity in a way faster than logarithmic growth, $z_3\rightarrow z_3\exp[l]$. As a result, the three energy scales should renormalize to infinity, as mentioned at the beginning of this subsection. This justifies our assumption that the probing energy scale is always much smaller than $\Lambda_{\text{f}}$, $m_{\text{p}}$ and $v_3^2m_{\text{s}}$.
	
	The most interesting feature of our result is that the fermion's band structure is strongly dressed by the soft photons. In particular, the velocity in $z$-direction, $v_3$, renormalizes to zero. Such result matches the qualitative expectation that the soft photons are essentially an impedance of fermion's motion in $z$-direction. One may wonder about the non-analyticity of the RG flow equations Eq.~\eqref{eq:EffVelocity} and \eqref{eq:EffCoupling} at $v_3=0$. This is because the momentum shell at the cutoff $\Lambda^2=q_0^2+q_1^2+q_2^2+v_3^2q_3^2$, as a surface in momentum space, has different topology for $v_3\neq 0$ (ellipse) and $v_3=0$ (infinite cylinder). The non-analyticity in $v_3$ also makes the velocity term \textit{dangerously irrelevant} [\onlinecite{cardy1996scaling}], and the limit $v_3\to 0$ should always be taken with caution, even if one is primarily interested in the fixed point properties. 

	\subsection{Photon Emission Rate and Fermion Lifetime}

	The second effect is related to the spontaneous emission of photons and correspondingly the fermion lifetime. As schematically illustrated in Fig.~\ref{fig:MeanFieldDispersion}(c), when moving in $z$-direction, fermions can spontaneously emit the soft photons. A comment on the definition of the soft (slow) and fast photon is due. Qualitatively, the criteria amounts to the comparison between the photon's phase velocity $c_{\text{phase}}$ and the fermion's group velocity $v_{\text{group}}$ [\onlinecite{schwinger1998classical}]. When the photon's phase velocity is small $c_{\text{phase}}<v_{\text{group}}$, the photons are soft and slow and can be emitted spontaneously. In the opposite limit, the photons are fast and cannot be emitted. A careful analysis of energy-momentum conservation generally gives a stricter kinematic constraint on the emitted photons.
	
	The one-photon emission rate can be calculated from the imaginary part of the fermion self-energy, Fig.~\ref{fig:OneLoopDiagrams}(a), by the optical theorem [\onlinecite{weinberg1995quantum},\onlinecite{peskin2018introduction},\onlinecite{kamenev2011field}]. When the external fermions are put on mass-shell, the result of the calculation is fully independent of gauge choice, according to the Ward identity [\onlinecite{weinberg1995quantum},\onlinecite{peskin2018introduction}]. Thus, the imaginary part of the on-mass-shell fermion self-energy is considered in this subsection, $\tau^{-1}=2\text{Im }\text{tr} \Sigma(k)\hat{P}_{\pm}(k)|_{k_0=\pm E(\boldsymbol{k})}$, where $\hat{P}_{\pm}(k)$ is the projection operator onto the positive (negative) energy fermion bands.
	
	\begin{figure}[t]
	    \centering
	    \includegraphics[scale = 0.425]{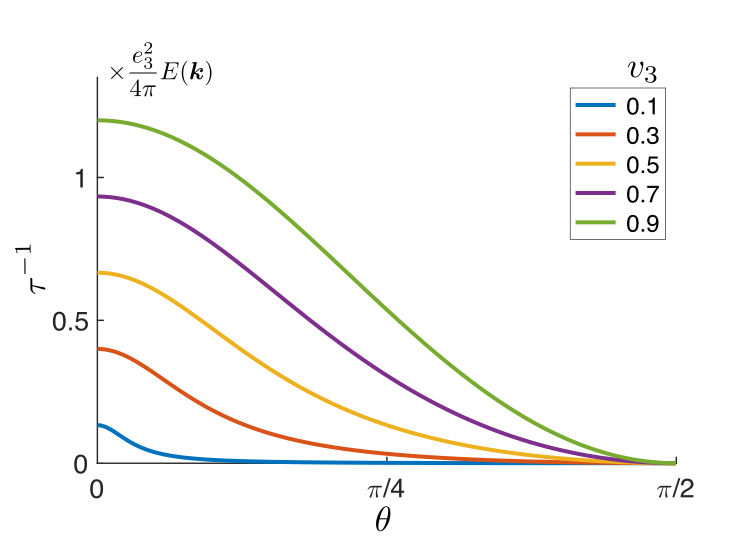}
	    \caption{Inverse fermion lifetime versus the direction of fermion momentum at various values of velocity $v_3$. Fermion has a momentum in the $xz$-plane $\boldsymbol{k}=|\boldsymbol{k}|(\sin\theta,0,\cos\theta)$ and energy $E(\boldsymbol{k})=|\boldsymbol{k}|\sqrt{\sin^2\theta+v_3^2\cos^2\theta}$. Here, the fermion energy is small compared to the typical energy scales in the system.}
	    \label{fig:FermionLifetime}
	\end{figure}
	
	Given the rotation symmetry in the $xy$-plane, we can assume the external fermion has a momentum in the $xz$-plane without loss of generality. Namely, the momentum of the external fermion is $\boldsymbol{k}=|\boldsymbol{k}|(\sin\theta,0,\cos\theta)$. To leading order in the coupling constant $e_3^2$, we found that the one-photon emission rate or equivalently the inverse fermion lifetime is given by:
	\begin{equation}
	    \tau^{-1}=\frac{e_3^2}{4\pi}E(\boldsymbol{k})\times \frac{4}{3}\frac{|v_3|^3\cos^2\theta}{\sin^2\theta+v_3^2\cos^2\theta}
	    \label{eq:FermionLifetime}
	\end{equation}
	It is proportional to the fine structure constant $\frac{e_3^2}{4\pi}$ and the fermion's energy $E(\boldsymbol{k})=|\boldsymbol{k}|\sqrt{\sin^2\theta+v_3^2\cos^2\theta}$. In addition, the fermion lifetime also depends on fermion's velocity and the direction of momentum.
	
	As plotted in Fig.~\ref{fig:FermionLifetime}, the inverse of the fermion lifetime increase with increasing velocity $v_3$. This matches the expectation of Cherenkov radiation [\onlinecite{schwinger1998classical}]. The inverse lifetime also depends on the direction of fermion's momentum. Indeed, the inverse lifetime is maximum when the fermion moves in the $z$-direction ($\theta=0$). Meanwhile, the fermions do not emit photons when moving in the $xy$-plane ($\theta=\pi/2$).
	
	At weak coupling, $e_3^2/4\pi\ll 1$, the fermions are indeed under-damped. The presumption of weak damping in the RG analysis is justified. The parameters in Eq.~(\ref{eq:FermionLifetime}) can be thought of as the renormalized ones. The inverse of the fermion life time is proportional to its energy, $\tau^{-1}\propto E(\boldsymbol{k})$. Meanwhile, the coefficient of proportionality approaches zero logarithmically upon approaching the IR limit.
	
	In addition, when the coupling strength is moderate $e_3^2/4\pi\lesssim 1$, the inverse fermion lifetime is still proportional to the energy. The ratio between the two quantities may not be small, $[\tau E(\boldsymbol{k})]^{-1}\lesssim\mathcal{O}(1)$. This is suggesting that the fermions are significantly damped at the moderate electromagnetic coupling. Accordingly, the fermionic quasiparticles are not sharply defined. We expect the RG still play some role in this regime. Then the coupling constants flow to a smaller value. Below certain scale, the fermionic excitation eventually becomes weakly or even non-interacting and under-damped. Therefore, the emergent WSM represents an unconventional quantum liquid. Over a wide energy window, it behaves like a marginal Fermi liquid and lacks sharply defined quasiparticles [\onlinecite{PhysRevLett.63.1996}]. 
	
	At strong coupling, the fermionic quasiparticles are over-damped. Then, perhaps a hydrodynamic description for such chiral plasma is in order, which we do not pursue in great detail here. 
	
	\section{Conclusion and Discussions}
	\label{sec:Conclusion}

	To conclude, we investigate the dynamical effect of the chiral anomaly, through the question of what is the infrared description of an emergent Weyl semimetal. We found that the emergent WSM represents an unconventional quantum liquid. The anomaly term guarantees the softness of the emergent photons. Due to the soft nature of the emergent U(1) gauge field, the IR behavior of the emergent WSM is quite distinct from the noninteracting WSM in two aspects. 
	
	First, the fermions get dressed by the soft photons significantly. The fermion velocity in the $z$-direction renormalizes to zero. In the IR limit, both fermions and gauge fields show an asymptotic 2D dispersion. Meanwhile, the emergent fine structure constant renormalizes to zero as well, leading to a decoupled IR limit. 
	
	Second, the fermions acquire a finite lifetime through spontaneous emission of soft photons. The inverse of fermion lifetime is found to be proportional to the fine structure constant and, more interestingly, the fermion energy. This is suggesting that the emergent WSM behaves like a marginal Fermi liquid at finite energy. However, due to the RG effect, the emergent WSM has a well-controlled (non-interacting) IR limit.

	The reported features of an emergent WSM is potentially detectable through specific heat measurement [\onlinecite{PhysRevLett.102.176404},\onlinecite{PhysRevLett.118.246601}]. The specific heat of fermions scales with temperature $T$ as $C_{\text{f}}\sim T^3/|v_3|$. At low temperature $T\ll m_{\text{s}}$, the contribution from the soft photons has a different power, $C_{\text{sp}}\sim m^{1/2}_{\text{s}}T^{5/2}$. Therefore, it's possible to single out the fermionic contribution to the specific heat [\onlinecite{kittel1996introduction}]. The measurement of fermionic specific heat reveals the velocity renormalization effect. In particular, $C_{\text{f}}/T^3\sim 1/|v_3|$ should show a logarithmic increase upon lowering the temperature. Notice that the acoustic phonon in 3D also has a specific heat of cubic temperature dependence, $C_{\text{ap}}\sim T^3$. We expect the renormalization effect on phonon dispersion is limited. Hence, $C_{\text{ap}}/ T^3$ is basically temperature independent. Therefore, after singling out the emergent photons, an increase in $(C_{\text{f}}+C_{\text{ap}})/T^3$ with lowering the temperature should reflect the renormalization effect of reducing the fermion velocity $v_3$ in the IR limit.

	
	\acknowledgements
	
	We are grateful to Sung-Sik Lee, Ruochen Ma, Alex Kamenev, Adam Nahum and Jinmin Yi for the valuable discussions. We acknowledge support from the Natural Sciences and Engineering Research Council (NSERC) of Canada. AAB was also supported by Center for Advancement of Topological Semimetals, an Energy Frontier Research Center funded by the U.S. Department of Energy Office of Science, Office of Basic Energy Sciences, through the Ames Laboratory under contract DE-AC02-07CH11358. Research at Perimeter Institute is supported in part by the Government of Canada through the Department of Innovation, Science and Economic Development and by the Province of Ontario through the Ministry of Economic Development, Job Creation and Trade.

	\bibliography{GaugedWSMReference}

\end{document}